\documentclass[aps,pra,superscriptaddress,twocolumn]{revtex4} 
\usepackage{subfigure}
\usepackage{amsmath}
\usepackage{amsfonts}
\usepackage{bm}
\usepackage[pdftex]{color,graphicx} 
\usepackage{gensymb}
\usepackage{amssymb}
\usepackage{textcomp}
\usepackage{subfigure}
\usepackage{bm}
\usepackage{color}
\usepackage{stmaryrd}   
\usepackage{siunitx}

\newcommand{\nn}{\nonumber}

\def\a{\alpha}
\def\b{\beta}

\def\d{\delta}

\def\th{\theta}

\def\r{\rho}

\def\s{\sigma}

\def\p{\psi}
\def\w{\omega}

\def\P{\Psi}

\def\<{\langle}
\def\>{\rangle}

\begin{document}

	\title{Entangling Bosons through Particle Indistinguishability and Spatial Overlap}

\author{Mariana R. Barros}
\thanks{These authors contributed equally.}
\affiliation{Center for Quantum Information, Korea Institute of Science and Technology (KIST), Seoul, 02792, Republic of Korea}

\author{Seungbeom Chin}
\thanks{These authors contributed equally.}
\affiliation{ Department of Electrical and Computer Engineering, Sungkyunkwan University, Suwon 16419, Republic of Korea}

\author{Tanumoy Pramanik}
\thanks{These authors contributed equally.}
\affiliation{Center for Quantum Information, Korea Institute of Science and Technology (KIST), Seoul, 02792, Republic of Korea}	
		
\author{Hyang-Tag Lim}
\affiliation{Center for Quantum Information, Korea Institute of Science and Technology (KIST), Seoul, 02792, Republic of Korea}
	
\author{Young-Wook Cho}
\affiliation{Center for Quantum Information, Korea Institute of Science and Technology (KIST), Seoul, 02792, Republic of Korea}

\author{Joonsuk Huh}%
\email{joonsukhuh@gmail.com}
\affiliation{Department of Chemistry, Sungkyunkwan University, Suwon 16419, Korea}
\affiliation{SKKU Advanced Institute of Nanotechnology (SAINT), Sungkyunkwan University, Suwon 16419, Republic of Korea}

\author{Yong-Su Kim}
\email{yong-su.kim@kist.re.kr}
\affiliation{Center for Quantum Information, Korea Institute of Science and Technology (KIST), Seoul, 02792, Republic of Korea}
\affiliation{Division of Nano \& Information Technology, KIST School, Korea University of Science and Technology, Seoul 02792, Republic of Korea}

\date{\today}
	
	\begin{abstract}
\noindent
Particle identity and entanglement are two fundamental quantum properties that work as major resources for various quantum information tasks. However, it is still a challenging problem to understand the correlation of the two properties in the same system.
While recent theoretical studies have shown that the spatial overlap between identical particles is necessary for nontrivial entanglement, the exact role of particle indistinguishability in the entanglement of identical particles has never been analyzed quantitatively before.
Here, we theoretically and experimentally investigate the behavior of entanglement between two bosons as spatial overlap and indistinguishability simultaneously vary. 
The theoretical computation of  entanglement for generic two bosons with pseudospins is verified experimentally in a photonic system.
Our results show that the amount of entanglement is a monotonically increasing function of both quantities.
We expect that our work provides an insight into deciphering the role of the entanglement in quantum networks that consist of identical particles.
	\end{abstract}
	
	
	\maketitle

	\paragraph*{Introduction.---}
Entanglement is one of the most significant quantum features, playing a principal role in the study of  both quantum foundation and quantum information. A  given entangled system cannot be characterized as a simple collection of individual subsystems, even when they are space-like separated.~\cite{einstein1935can, bell2004speakable}. It is also a crucial resource of various quantum tasks including quantum computation~\cite{ladd2010}, quantum communication~\cite{ekert1991quantum,bennett1993teleporting}, and quantum sensing~\cite{lugiato2002quantum}. 
	
   However, while the entanglement between non-identical particles is mathematically well-described as the superposition of different multipartite states~\cite{schrodinger1935mathematical}, its formalism cannot be directly applied to the case of identical particles. A set of identical particles has  \emph{the exchange symmetry}~\cite{dirac1981principles}, which makes the Hilbert space of the particles non-factorizable, and thus, every state of identical particles becomes a superposed multipartite state in the first quantization language. Hence, one can say that any particles that are identical to each other are entangled even when they are prepared in worlds apart~\cite{peres2006quantum}, which seems implausible. This is called \emph{the non-factorization problem} of identical particles.
   
   To solve this puzzle, we need to discard unphysical entanglement  from the physical one. Such unphysical entanglement is merely mathematical and, therefore, it cannot be extracted by any physical setup~\cite{ schliemann2001j,schliemann2001ja, ghirardi1977gc,ghirardi2002entanglement, eckert2002k, paskauskas2001r, zanardi2002p,shi2003shi,barnum2004subsystem, barnum2005generalization}. Indeed, when we say a multipartite system has nontrivial entanglement, the  entanglement can be detected by registers that are distinguishable to each other~\cite{wiseman2003entanglement,killoran2014extracting}. There are several approaches to formally define the entanglement of identical particles to overcome the non-factorization problem of identical particles. The recent crucial formalisms include algebraic approaches \cite{balachandran2013entanglement, balachandran2013algebraic},  the second quantization approach~\cite{benatti2011entanglement,benatti2012bipartite,lourencco2019entanglement}, no-labeling approach~\cite{franco2016quantum,giuseppe2018,compagno2018dealing}, and the first quantization approach (1QL)~\cite{chin2019entanglement}. All these formalisms provide criteria to distinguish the mathematical entanglement from physically extractable entanglement by using exchange symmetry. The mathematical connection among the aforementioned approaches is explained in Ref.~\cite{chin2019reduced} from the viewpoint of the partial trace. 

 One of the imperative queries closely related to the above studies on the entanglement of identical particles is: \emph{under what conditions can a set of identical particles carry physical entanglement?} Answering this question not only provides an insight to understand the fundamental property of entanglement, but also practically helps to construct quantum systems that employ identical particles as information processing resource~\cite{monroe2002quantum}. It is well-known that such set is able to possess entanglement only when the particles spatially overlap~\cite{paunkovic2004role,tichy2013entanglement,giuseppe2018}, or more rigorously speaking when non-zero spatial coherence between particles and detectors exists~\cite{chin2019entanglement}.  Hence, one can surmise that the particle indistinguishability and spatial overlap are two key factors for identical particles to be entangled. On the other hand, while theoretical studies on the quantitative relation of entanglement to spatial overlap alone have been executed before~\cite{tichy2013entanglement,giuseppe2018,chin2019entanglement}, there has been no attempt to examine the effect of both indistinguishability and spatial overlap to the entanglement of identical particles at the same time either theoretically or experimentally. 
 
In this Letter, we theoretically predict the quantitative relation of entanglement between two bosons with particle indistinguishability and spatial overlap using recently proposed partial trace technique~\cite{franco2016quantum,giuseppe2018,compagno2018dealing,chin2019entanglement}, which is then experimentally verified using photons. We show that the amount of entanglement, which is quantified as concurrence, is a monotonically increasing function of both indistinguishability and spatial overlap. The obtained data confirm that \emph{two kinds of quantum properties, i.e., particle indistinguishability and spatial overlap, cooperate to generate entanglement.}

\noindent\paragraph*{Theory.---} 

This section presents a theoretical analysis of the effect of particle indistinguishability and spatial overlap on entanglement generation. We first introduce a formalism for calculating the entanglement between two bosons in the generic system. Then, we derive a formula for concurrence to investigate the behavior of entanglement with the variation of the distinguishability and spatial overlap. 


Suppose that we have a system of two bosons that are in the states ($\P_A$, $\P_B$). Then, from the exchange symmetry of bosons, the total state $|\P\>$ can be written as 
\begin{align}\label{nolabeling}
 |\P\> = 
 |\P_A,\P_B\>
\end{align}
with the symmetry relation
\begin{align}
   |\P_A,\P_B\> = |\P_B,\P_A\>.
\end{align}
The transition relations between two different wave functions are given by~\cite{franco2016quantum}
\begin{align}\label{projection}
&\< \P_C, \P_D|\P_A,\P_B\> = \<\P_C|\P_A\>\<\P_D|\P_B\> \nn \\
&\quad\qquad\qquad\qquad \quad  + \<\P_C|\P_B\>\<\P_D|\P_A\>,\nn \\
&\<\P_C|\P_A,\P_B\> = \frac{1}{\sqrt{2}}\Big[\<\P_C|\P_A\>|\P_B\> +\<\P_C|\P_B\>|\P_A\>\Big].
\end{align} 
The above relations can be derived explicitly using the first quantization language (see Supplemental Materials). 

Since our goal is to clarify the role of indistinguishability and spatial overlap in the generation of entanglement, we need three degrees of freedom to present the particle states, i.e., internal pseudospin (which will be detected at the output modes), distinguishability function, and spatial distribution function. Hence, $\P_i$ ($i=A,B$) are now specified as $\P_i = (\p_i, s_i, \phi_i)$ where $\p_i$ is the spatial wave function, $s_i$ is the spin value, and $\phi_i$ is the particle distinguishability. Considering that the result becomes trivial for $s_A=s_B$, we restrict our case into $(s_A,s_B) = (\uparrow,\downarrow)$. 
The total state of two bosons is now presented as
\begin{align}\label{wavefcn;gen}
 |\P\> = |(\p_A, \uparrow, \phi_A), (\p_B,\downarrow,\phi_B)\>. 
\end{align}
Note that at this point one cannot tell about the entanglement of $|\P\>$, since Eq.~\eqref{wavefcn;gen} says nothing about the spatial relation of particles to detectors. On the other hand, the entanglement of identical particles is affected by the measurement setup~\cite{tichy2013entanglement, chin2019entanglement}. 

\begin{figure}[t]
	\includegraphics[width=3in]{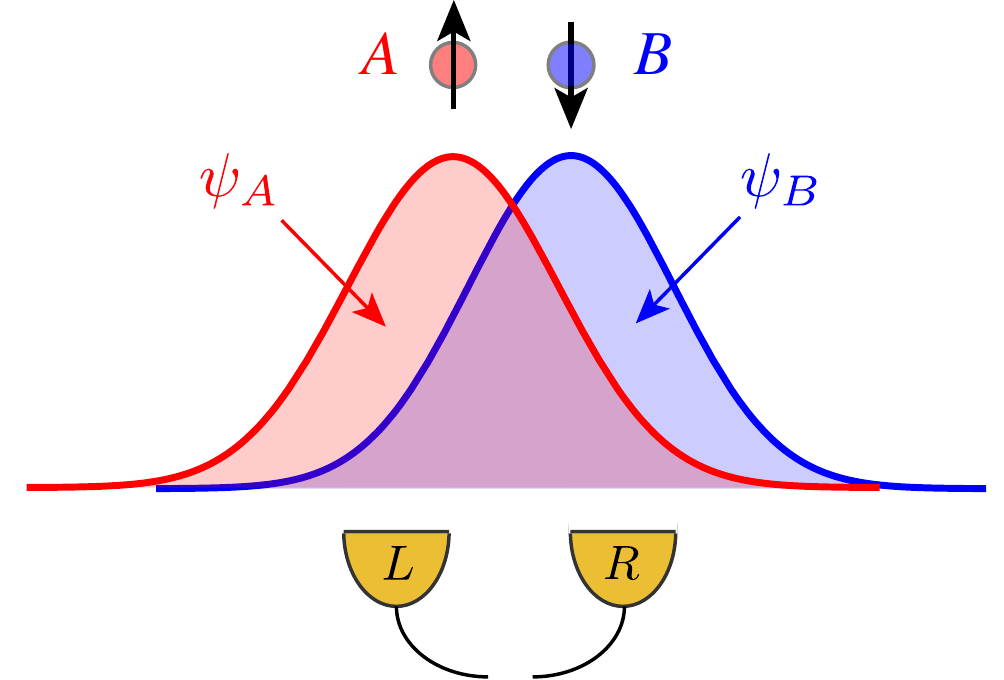}
	\caption{Generic setup for detecting entanglement of two identical particles: The spatially overlapped particles have internal spins ($\uparrow$ and $\downarrow$) and distinguishability (represented as two different colors, red and blue, in the figure). Two detectors located at $L$ and $R$ indicate the spins of the particles.}	\label{fig:concept_general}
\end{figure}

In order for a multipartite system to have physical entanglement, the subsystems should be individually accessible by observers~\cite{dalton2017quantum}. 
In addition, we consider the entanglement of general bosons, which conserve particle number superselection rule (SSR)~\cite{wickwigner}. One of such kinds of entanglement is the \emph{entanglement of particles} introduced in Ref.~\cite{wiseman2003entanglement}. Hence, our setup needs two detectors $L$ and $R$ that are located in distinctive places (spatially separated, i.e., $\<L|R\>=0$) that can measure the superposition of wavefunctions with the same particle number distribution but different internal states.

Figure~\ref{fig:concept_general} is a conceptual diagram to extract physical entanglement from two bosons.
Since we are interested in the case only when two particles are detected at two observers $L$ and $R$,
the spatial relations of the particles to the detectors are expressed in the most general form as
\begin{align}
|\p_A\> = \a_l|L\> + \a_r|R\>, \quad |\p_B\> = \b_l|L\> + \b_r|R\>
\end{align} 
where $\{\a_r,\a_l,\b_r,\b_l\}$ are complex numbers and satisfy $|\a_l|^2 + |\a_r|^2 =|\b_l|^2 + |\b_r|^2 =1$. Then, Eq.~\eqref{wavefcn;gen} can be rewritten in terms of the detector basis $\{|L\>,|R\>\}$ as 
\begin{align}\label{afterdetection}
  |\P\> &= \a_l\b_l |(L,\uparrow,\phi_A),(L,\downarrow,\phi_B)\> \nn \\
  & + \a_l\b_r|(L,\uparrow,\phi_A),(R,\downarrow,\phi_B)\> \nn \\
  & + \a_r\b_l|(L,\downarrow,\phi_B),(R,\uparrow,\phi_A)\> \nn \\
  & + \a_r\b_r |(R,\uparrow,\phi_A),(R,\downarrow,\phi_B)\>.
\end{align}

From the definition of the particles entanglement (See Supplemental Materials for details), the only terms of Eq.~\eqref{afterdetection} that participate in the entanglement are (in an un-normalized form) given by
\begin{align}\label{afterpostselection}
  \a_l\b_r|(L,\uparrow,\phi_A),(R,\downarrow,\phi_B)\> + \a_r\b_l|(L,\downarrow,\phi_B),(R,\uparrow,\phi_A)\>.
\end{align}

Since we measure the spin states of the single particles in the two detectors, the quantum state $\r $ expected to be measured is the one of which the distinguishability function is traced out. Hence, $\r$ is given by
\begin{align}\label{density}
 \r
  &= |\a_l\b_r|^2 |(L,\uparrow),(R,\downarrow)\>\<(L,\uparrow),(R,\downarrow)|  \nn \\
  &+ |\a_r\b_l|^2 |(L,\downarrow),(R,\uparrow)\>\<(L,\downarrow),(R,\uparrow)|  \nn \\
  &+ \a_l\b_r\a_r^*\b_l^*|\<\phi_A|\phi_B\>|^2 |(L,\uparrow),(R,\downarrow)\>\<(L,\downarrow),(R,\uparrow)| \nn \\ &+\a_l^*\b_r^*\a_r\b_l|\<\phi_A|\phi_B\>|^2|(L,\downarrow),(R,\uparrow)\>\<(L,\uparrow),(R,\downarrow)|.
\end{align}
The concurrence of the state $\r$, which quantifies the amount of entanglement, is computed as~\cite{hill1997entanglement,wootters1998entanglement}
\begin{align}\label{concurrence}
C= 4|\a_l\a_r\b_l\b_l|\cdot|\<\phi_A|\phi_B\>|^2.
\end{align}
Here $4|\a_l\a_r\b_l\b_l|$ and  $|\<\phi_A|\phi_B\>|^2$ represent the spatial overlap and the particle indistinguishability, respectively. The derivation of Eqs.~\eqref{density} and \eqref{concurrence} is explained in the Supplemental Materials. 

Equation~\eqref{concurrence} shows  that the concurrence is determined by three variables, $|\a_l|$, $|\b_l|$, and $|\<\phi_A|\phi_B\>|$.
The particles are unentangled when they are completely distinguishable, i.e., $\<\phi_1|\phi_2\>=0$, or when one of the particles is detected at only one side, i.e., one of $\{\a_l,\a_r,\b_l, \b_r\}$ is zero. On the other hand, the particles are maximally entangled when the indistinguishability and spatial overlap are both maximal, i.e., $|\<\phi_1|\phi_2\>|=1$ and  $|\a_l| = |\b_l|=\frac{1}{\sqrt{2}}$. 

\begin{figure}[b]
	\centering 
	\includegraphics[width=3.4in]{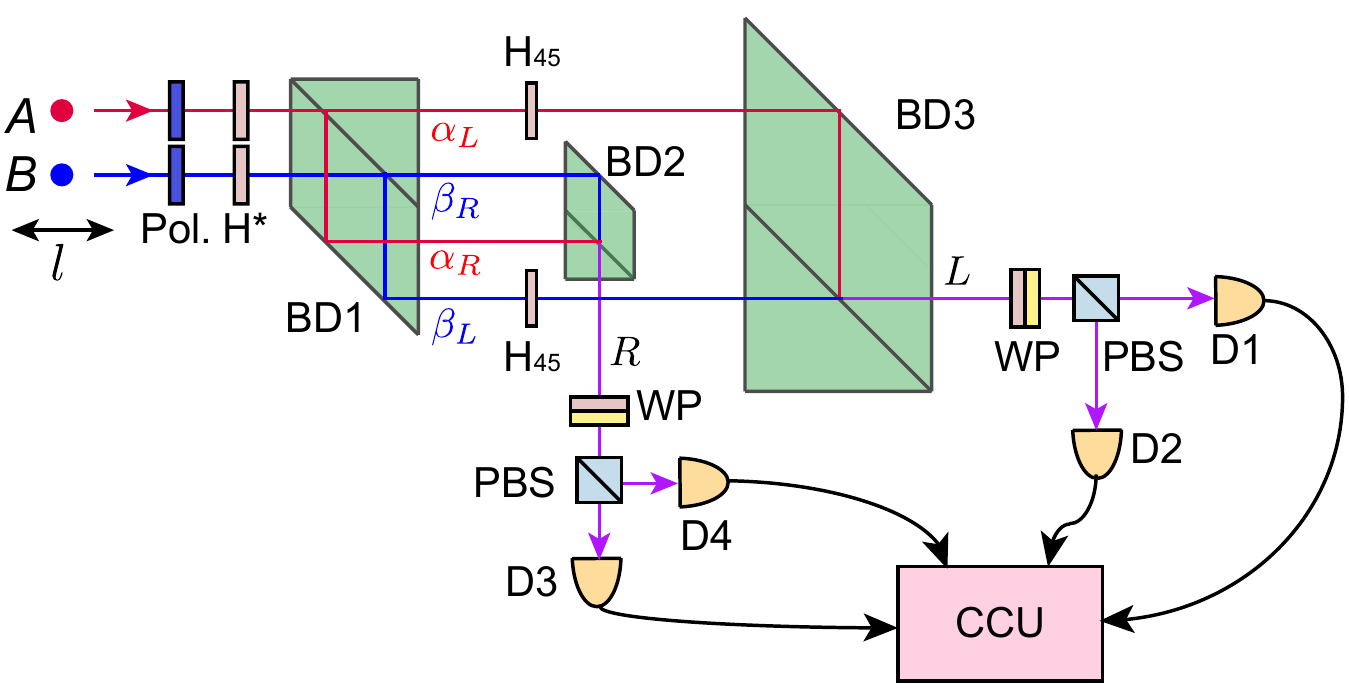}
	\caption{The experimental setup to entangling photons via their spatial overlap and indistinguishability. Pol. : polarizer, H$^*$ : half waveplate at $\theta$, BD : beam displacer, H$_{45}$ : half waveplate at 45$\degree$, WP : waveplates, PBS : polarizing beamsplitter, D1-D4 : single photon detectors, CCU : coincidence counting unit.}
	\label{experiment} 
\end{figure} 


\paragraph*{Experiment.---}

In an optical system, photons are the identical particles and the photon polarization $\{|H\rangle,|V\rangle\}$ compose the spin $\{|\downarrow\rangle,|\uparrow\rangle\}$, where $|H\rangle$ and $|V\rangle$ denote the horizontal and vertical polarization states, respectively. Particle distinguishability can be controlled by other degrees of freedom such as momenta, spectra, and temporal modes. Here, it is controlled by the temporal modes of single photons.

The identical photon pairs, centered at 780~nm, were generated at a type-II BBO crystal via spontaneous parametric down conversion pumped by a femtosecond pulse at 390~nm~\cite{wang2016experimental,pramanik2019revealing,kim18} . To guarantee the spectral and spatial indistinguishability, two photons were filtered by 3~nm bandpass interference filters (IF) at 780~nm, then collected by single-mode fibers (SMF). Next, they were sent to the experimental setup shown in Fig.~\ref{experiment} in order to investigate entanglement behavior via particle indistinguishability and spatial overlap. 

The initial polarization states of the incoming photons $A$ and $B$ were configured by polarizers (Pol.) and half waveplates (HWP, H$^*$). Then, its probability amplitudes were divided into two spatial modes by a beam displacer (BD1) which transmits and reflects horizontal and vertical polarization states, respectively. The HWP at $45\degree$ (${\rm H}_{45}$) at spatial modes $\alpha_L$ and $\beta_L$ converted the polarization states from horizontal to vertical and vice versa. Therefore, the polarization states of photon $A$ and $B$ became vertical and horizontal, respectively, regardless of the spatial modes. Subsequently, the spatial modes $\alpha_R$ and $\beta_R$ ($\alpha_L$ and $\beta_L$) were combined by BD2 (BD3) which corresponds to the spatial wavefunction overlap. We measured the two-qubit polarization state at outputs of BD2 and BD3 ($R$ and $L$) using waveplates (WP), polarizing beamsplitters (PBS) and a coincidence counting unit (CCU)~\cite{ccu}. 

The experimental setup faithfully demonstrates the conceptual diagram of Fig.~\ref{fig:concept_general}. The degree of spatial overlap at $L$ and $R$ could be adjusted by changing the initial polarization states. By using $|H\rangle$ transmitting polarizers and HWPs at $\theta$, the initial polarization state becomes $\cos2\theta|H\rangle+\sin2\theta|V\rangle$. Note that we set the same initial polarization states for the photons $A$ and $B$ in order to investigate the symmetrical spatial overlap case. After BD1, the coefficients of polarization state are transferred to the probability amplitudes of spatial modes, so $\alpha_L=\beta_R=\sin2\theta$ and $\alpha_R=\beta_L=\cos2\theta$. Therefore, one can find the degree of spatial overlap becomes
\begin{equation}
4|\alpha_L\alpha_R\beta_L\beta_R|=\sin^2{4\theta}.
\label{spatial_overlap}
\end{equation}
Hence, the degree of distinguishability can be tuned by changing one of the photons arrival time at the BDs. This can be done by scanning one of the input photons, see the optical delay $l$ of Fig.~\ref{experiment}. Lastly, the successful case in which only one particle is found at $L$ and the other at $R$ can be post-selected by coincidences between $L$ and $R$. Note that one can trace out the indistinguishability wavefunction $\phi_i$ if the coincidence window is larger than the timing difference between $A$ and $B$, so the timing information is not registered. Assuming that the degree of distinguishability is presented as a Gaussian function with the center $l=0$, Eq.~\eqref{concurrence} can be rewritten with the variables of our optical system as
 \begin{align}\label{concurrence;optical}
     C&= \sin^2{4\th}\exp\left[-\frac{ l^2}{2\sigma^2} \right].
 \end{align}
The derivation of Eq.~\eqref{concurrence;optical} is explained in detail in the Supplemental Materials.

\begin{figure}[t]
\centering
\includegraphics[width=3in]{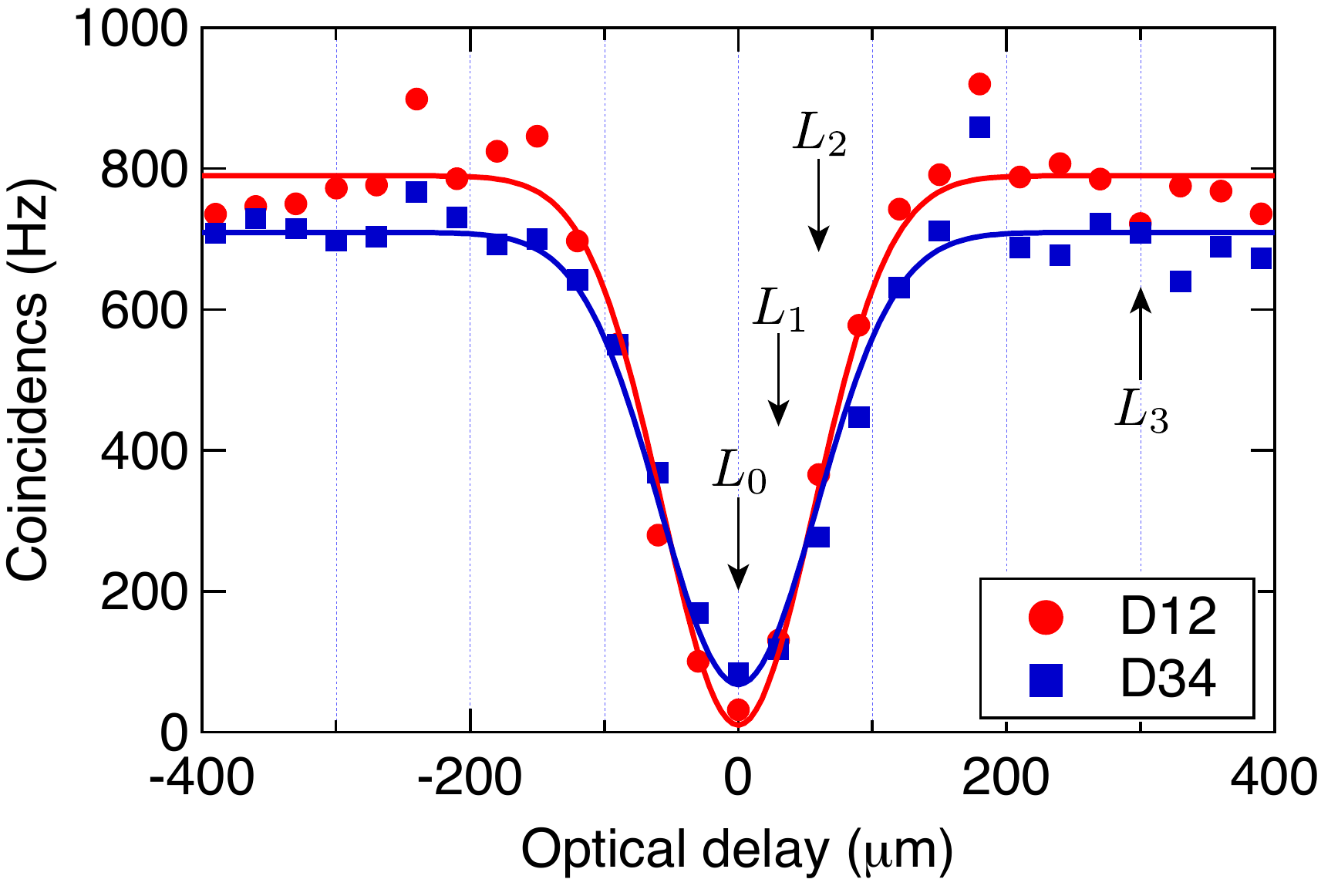}
\caption{Coincidence counts with respect to scanning the optical delay. A HOM dip can be observed for the coincidences D12 (experimental data and fitting curve in red) and D34 (experimental data and fitting curve in blue), the visibilities were $V=0.99\pm 0.04$ and  $V=0.91\pm 0.05$, respectively.}
	\label{HOM} 
\end{figure}

The distinguishability introduced by the optical delay $l$ can be measured by the Hong-Ou-Mandel (HOM) interference at BD2 and BD3. Figure~\ref{HOM} presents two-fold coincidences of D12 and D34 with respect to the optical delay $l$ where D{\it ij} stands for the coincidences between D{\it i} and D{\it j}. Therefore, the input polarization state was prepared as $|++\rangle$ where $|+\rangle=\frac{1}{\sqrt{2}}(|H\rangle+|V\rangle)$ in order to measure the HOM interference at BD2 and BD3, simultaneously. The HOM dip at $l=0$ implies that the two photons $A$ and $B$ are indistinguishable. The HOM interferences of D12 and D34, which can be fitted with the Gaussian function with the full width at half maximum (FWHM), are $132\pm4~\SI{}{\micro\metre}$ and $137\pm4~\SI{}{\micro\metre}$, respectively. These widths coincide with the coherence length of the photon pair which is determined by the bandwidth of IF. The HOM visibility, defined by the relative depth of the dip to the non-interfering counts, was measured as $V=0.99\pm 0.04$ and  $V=0.91\pm 0.05$ for D12 and D34, respectively. To investigate the role of particle indistinguishability for generation of entanglement, we have analyzed two-qubit polarization states at various optical delays of $L_0=\SI{0}{\micro\metre}$, $L_1=\SI{30}{\micro\metre}$, $L_2=\SI{60}{\micro\metre}$, and $L_3=\SI{300}{\micro\metre}$.

\begin{figure}[t]
	\includegraphics[width=3in]{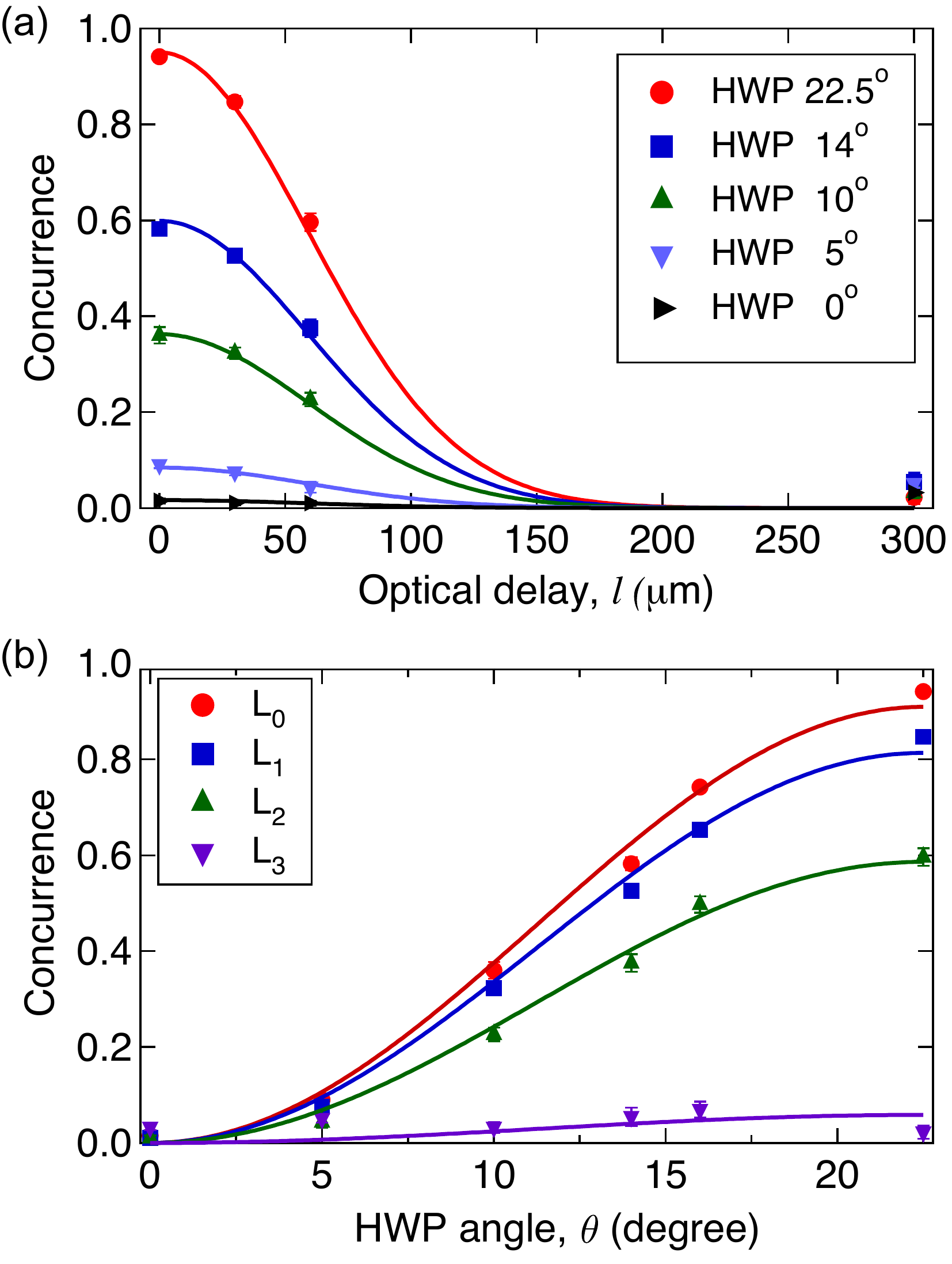}
	\caption{Concurrence with respect to (a) particle distinguishability, and (b) spatial overlap, respectively. The particle distinguishability is adjusted by the optical delay $l$, and the spatial overlap is controlled by the angle $\theta$ of the H$_\theta$. The error bars are obtained by performing 100 Monte Carlo simulation runs by taking into account the Poisson statistics in measured coincidence counts.}
	\label{data} 
\end{figure}

Figure~\ref{data} (a) presents concurrence as a function of optical delay $l$ at different HWP ${\rm H}^*$ angles $\theta$. It corresponds to the amount of entanglement with respective to particle distinguishability with fixed different degrees of spatial overlap. It clearly shows that concurrence decreases as the optical delay increases, or equivalently the distinguishability of particles increase. The experimental data are fitted with a Gaussian function with the FWHM of $140~\SI{}{\micro\metre}$, comparable to that of HOM interference. 

Figure~\ref{data} (b) represents concurrence as a function of the HWP ${\rm H}^*$ angle $\theta$ at different optical delays $l$. Note that $\theta=0\degree$ and $22.5\degree$ correspond to non-spatial overlap and complete spatial overlap cases, respectively. It clearly presents that the amount of entanglement increase as the spatial overlap increases. As expected from Eq.~(\ref{spatial_overlap}), the experimental data are fitted with $C=C_0\sin^2{4\theta}$, where $C_0$ corresponds to the maximum concurrence at a given optical delay $l$, and $\theta$ in radian.

\paragraph*{Conclusion.---}
In this work, we have investigated the quantitative relation of the entanglement between identical particles with particle indistinguishability and spatial overlap from the perspectives of theory and experiment. The entanglement concurrence of two oppositely polarized photons decreases with decreasing particle indistinguishability and spatial overlap, which fits in well with the theoretical predictions for the entanglement of identical particles in the generic boson system. 
Our analysis can be applied to various quantum information processing utilizing identical particles as media of information. For example, the computational difficulty of the boson sampling problem~\cite{aaronson2011computational} is known to have a close relation with the distinguishability~\cite{shchesnovich2015partial,tichy2015sampling} and the spatial distribution of photons~\cite{chin2019majorization}. We expect that our work provides an insight into understanding the role of the entanglement of identical particles in quantum information processing such as linear optical quantum computing.
In addition, our method can also be applied to fermions. They follow the \emph{parity SSR}~\cite{wickwigner,wick1970superselection}, which makes the entanglement of fermions more complicated  as well as intriguing~\cite{friis2016reasonable,gigena2017bipartite}.


\paragraph*{Acknowledgements.---} This work was supported by the National Research Foundation of Korea (NRF) (2019R1A2C2006381, 2019M3E4A1079777, 2019M3E4A1079666), and the KIST research program (2E29580). S.C. is supported by NRF funded by the Ministry of Education (NRF-2019R1I1A1A01059964). J.H is supported by NRF funded by the Ministry of Education, Science and Technology (NRF-2015R1A6A3A04059773) and the POSCO Science Fellowship of POSCO TJ Park Foundation.


%


	\pagebreak
	\begin{center}
		\textbf{\large Supplemental Materials}
	\end{center}
	\setcounter{page}{1}
	\makeatletter
	

\section{Derivation of transition relations between bosonic states}\label{1QL}
Here we derive the transition relations Eq.~\eqref{projection} using the first quantization language (1QL). 

In 1QL~\cite{chin2019entanglement}, the wavefunction of a particle with pseudo-label $a$ with a physical state $\P$ is described by $|\P\>_a$. Note that the particle pseud-labels are  mathematical artifacts that cannot be detected in principle. 
Then, the transition amplitude from $|\P\>_a$ to $|\Phi\>_b$ is given by
\begin{align}\label{transition}
 \<\Phi|_a\cdot |\P\>_b = \<\Phi|\P\>\d_{ab}.
\end{align}
The total state of $2$ bosons in states $\P_A$ and $\P_B$, $|\P_A,\P_B\>$ is written in the symmetrized form as 
\begin{align}
|\P_A,\P_B\> = \frac{1}{\sqrt{2}}\Big(|\P_A\>_1|\P_B\>_2 + |\P_B\>_1|\P_A\>_2 \Big),
\end{align}
and a one-particle subsystem with physical state $\P_C$ in the symmetrized form is given by
\begin{align} 
|\P_C\> = \frac{1}{\sqrt{2}} \Big(|\P_C\>_1 +|\P_C\>_2 \Big). 
\end{align} Here we just ignored the relative phase between $|\P_C\>_1$ and $|\P_C\>_2$ that is  an unphysical gauge~\cite{chin2019reduced}. 
Then the transition relations Eq.~\eqref{projection} are obtained directly from Eq.~\eqref{transition} as follows:
\begin{align}\label{proj1}
\< \P_C\P_D|\P_A,\P_B\> 
&=
\frac{1}{2}\Big(\<\P_C|_1\<\P_D|_2 + \<\P_D|_1\<\P_C|_2 \Big) \nn \\
& \qquad \cdot  \Big(|\P_A\>_1|\P_B\>_2 +|\P_B\>_1|\P_A\>_2 \Big) \nn \\
& =
\<\P_C|\P_A\>\<\P_D|\P_B\> \nn \\
&\quad  + \<\P_C|\P_B\>\<\P_D|\P_A\>,
\end{align}
\begin{align}
\<\P_C|\P_A,\P_B\>
&= \frac{1}{2}\Big(\<\P_C|_1+\<\P_C|_2\Big) \nn \\
& \qquad \cdot  \Big(|\P_A\>_1|\P_B\>_2 + |\P_B\>_1|\P_A\>_2 \Big) \nn \\
&= \frac{1}{\sqrt{2}}\Big( \<\P_C|\P_A\>|\P_B\> +\<\P_C|\P_B\>|\P_A\>\Big).
\end{align}
For the transition relations of arbitrary $N$ bosons (fermions), see Ref.~\cite{chin2019entanglement} (Ref.~\cite{chin2019reduced}). 

\section{Entangment of particles $E_P$ and concurrence $C$}\label{densitymatrix}

Massive bosons conserve 
particle-number conserving superselection rule (SSR)~\cite{wickwigner}, by which states with different particle numbers cannot superpose with each other. The entanglement of particles~\cite{wiseman2003entanglement}) is introduced to compute the entanglement of bipartite sytems that respect the particle number SSR, which is defined as
\begin{align}
    E_{P}(|\P\>) = \sum_{\vec{N}}p_{\vec{N}}E(|\P\>_{\vec{N}}),
    \label{entofpart}
\end{align}
where $\vec{N} = (n_L,n_R)$ is a possible number distribution vector ($n_R$ and $n_L$ is the number of particles in the right and left detector with $n_R + n_L=N$), $p_{\vec{N}}$ is the probability for a given state $|\P\>$ to have a number distribution $\vec{N}$, and $|\P\>_{\vec{N}}$ is the partial state of $|\P\>$ that has  $\vec{N}$. It is worth noting that the states of massless bosons such as photons are not restricted by the particle number SSR, hence an entanglement between different particle number distributions, i.e., mode entanglement, can also be considered.

For the case of Eq.~\eqref{afterdetection}, $|\P\>_{\vec{N} =(1,1)}$ is equal to Eq.~\eqref{afterpostselection}, which is the only part that contributes to $E_{P}(|\P\>)$. 
Since we are supposed to only measure the pseudospin states, the observed particle state is that which the distinguishability part of $|\P\>_{\vec{N} =(1,1)}$ is traced out. 
By defining the complete orthonormal basis of particle distinguishability as $\{|X_a\>\}$ so that 
\begin{align}\label{basis}
    \sum_{a}\<\phi|X_a\>\<X_a|\phi\>=1
\end{align}
holds for any $\phi,$ the state density matrix $\r$ is given by
\begin{align}
 \r 
  =& \sum_{a,b} \<(L,X_a),(R,X_b)|\P\>\<\P|(L,X_a),(R,X_b)\>. 
 \end{align} 
 And Eq.~\eqref{proj1} gives
 \begin{align}
    &\<(L,X_a),(R,X_b)|(L,\uparrow,\phi_1),(R,\downarrow,\phi_2) \> \nn \\
    &=\<X_a|\phi_1\>\<X_b|\phi_2\> |(L,\uparrow),(R,\downarrow)\>
 \end{align} and so on, by which 
 we obtain
 \begin{align}
  \r =& |\a_l\b_r|^2 |(L,\uparrow),(R,\downarrow)\>\<(L,\uparrow),(R,\downarrow)|  \nn \\
  &+ |\a_r\b_l|^2 |(L,\downarrow),(R,\uparrow)\>\<(L,\downarrow),(R,\uparrow)|  \nn \\
  &+ \a_l\b_r\a_r^*\b_l^*|\<\phi_1|\phi_2\>|^2 |(L,\uparrow),(R,\downarrow)\>\<(L,\downarrow),(R,\uparrow)| \nn \\ &+\a_l^*\b_r^*\a_r\b_l|\<\phi_1|\phi_2\>|^2|(L,\downarrow),(R,\uparrow)\>\<(L,\uparrow),(R,\downarrow)|.
\end{align}
Note that the last two terms of the above equation contribute to the non-trivial entanglement. Using the definition of concurrence for a mixed state given in Ref.~\cite{hill1997entanglement}, we have
Eq.~\eqref{concurrence}.

\begin{figure*}[t] 	
	\includegraphics[width=5in]{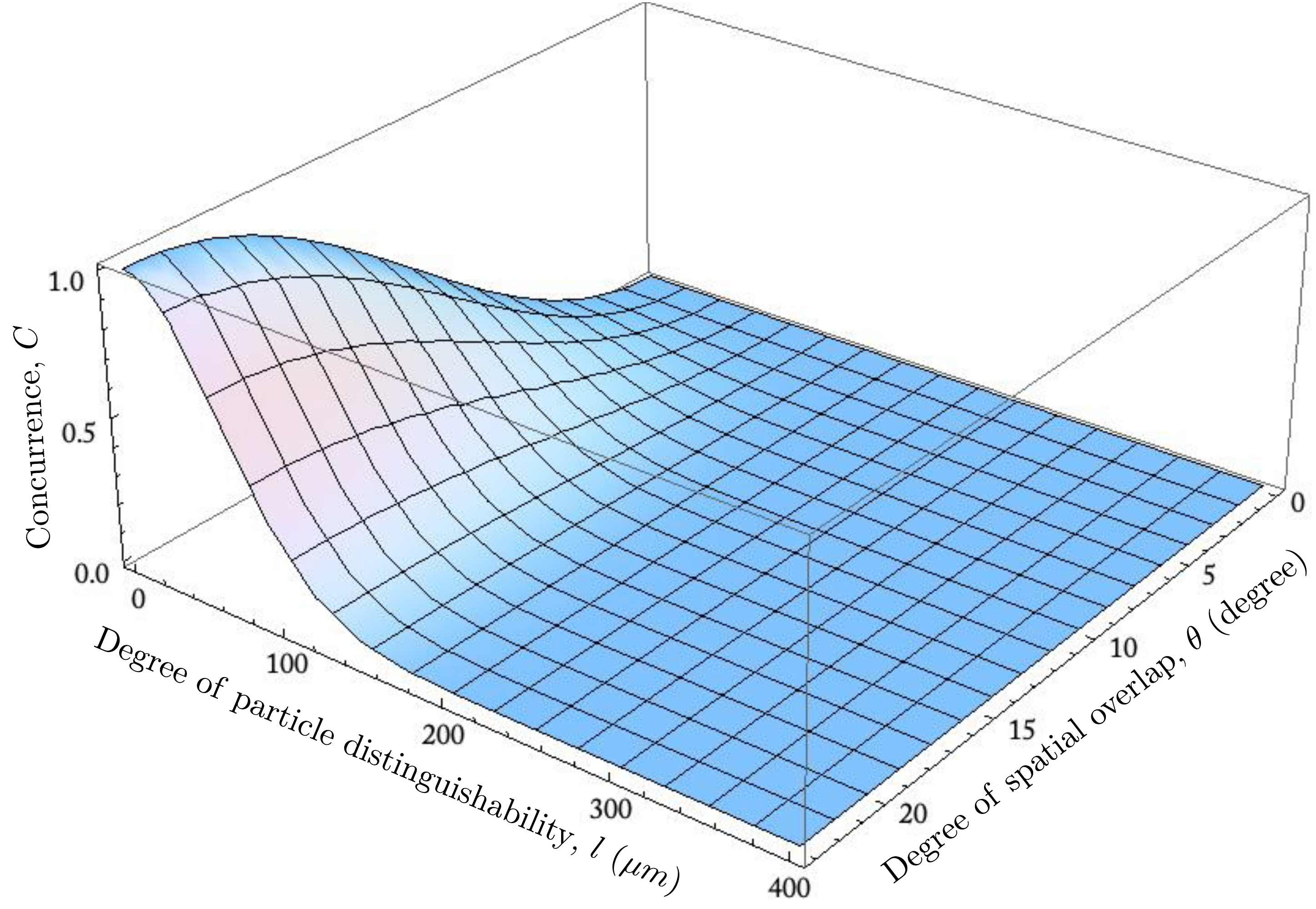}
    \caption{The 3-dimensional graph for the relation of concurrence $C$ with spatial overlap $\th$ and distinguishability $l$ in theory. $C$ is a sinusoidal function of $\th$ and a Gaussian fuction of $l$. $C$ has the maximal value $1$ when $\th=22.5$ and $l=0$.}
    \label{theorycurve}
\end{figure*}

To experimentally verify Eq.~\eqref{concurrence} in optics, the variables in the equation must be clarified as the optical ones.
In our scheme (Fig.~\ref{experiment}), the spatial overlap of two photons at $A$ and $B$ is controlled by changing the initial polarization states and the  distinguishability is determined by the optical delay between two photons $l$.

Denoting the arrival time of $A$ and $B$ as $t_A$ and $t_B$ (hence $l=|t_A-t_B|$), the total input state of two photons is given in a Gaussian form by
\begin{align}
 |\P\>= \int d\w_Ad\w_B\phi_A(\w_A)\phi_B(\w_B)|(A, H, w_A),(B,V,w_B)\>,
\end{align}
where 
\begin{align}\label{timedilation}
&\phi_A(\w_A) = \frac{1}{(2\pi\delta^2)^{\frac{1}{4}}}\exp\Big[i\w t_A-\frac{\w_A^2}{4\d^2}\Big], \nn \\
&\phi_B(\w_B) = \frac{1}{(2\pi\delta^2)^{\frac{1}{4}}}\exp\Big[i\w t_B -\frac{\w_B^2}{4\delta^2}\Big]. \nn \\
&\qquad\quad \qquad \qquad(\delta: \textrm{the spectral width})
\end{align} 

By substituting
\begin{align}
\<\phi_A|\phi_B\> &= \int d\w \phi_A^*(\w)\phi_B(\w) \nn \\
&= \exp\Big(-2\delta^2 (t_A-t_B)^2 \Big) \equiv \exp\Big(-2\delta^2 l^2 \Big)
\end{align} into Eq.~\eqref{concurrence},
and defining
\begin{align}
(|\a_l|, |\a_r|, |\b_l|, |\b_r|)=(\cos A,\sin A, \sin B, \cos B),
\end{align}
 we obtain the concurrence equation for optics
\begin{align}\label{concurrence;opt}
C&= \sin (2A)\sin(2B)\exp\Big(-\frac{ l^2}{2\sigma^2} \Big),
\end{align} where $\s$ is the standard deviation ($\s\equiv \frac{1}{2\d} $) of $C$.

 Since  $A$ and $B$ are set as equal to $2\th$ in our setup, the above equation is rewritten as Eq.~\eqref{concurrence;optical}. 
The relation of the concurrence $C$ with $\th$  and $l$ is displayed as a 3-dimensional curve in Figure~\ref{theorycurve}.


\end{document}